# The ReLiC– Recycling Linear e+e- Collider

Vladimir N Litvinenko [1,2], Nikhil Bachhawat[1], Maria Chamizo-Llatas[3], Yichao Jing[2,1], François Méot [2,1], Irina Petrushina[1], Thomas Roser[2]

[1] Department of Physics and Astronomy, Stony Brook University
[2] Collider-Accelerator Department, Brookhaven National Laboratory
[3] Physics Department, Brookhaven National Laboratory

*Status of Design: Concept*

**Introduction.** In this white paper we describe a concept of e+e- linear collider recycling both the used particles and the used beam energy – the ReLiC (Fig.1). The concept is based on segmenting superconducting (SRF) linear accelerators into sections divided by separators, where used (decelerating) beams are separated from colliding with accelerating beams by a combination of DC electric and magnetic fields. This design provides for undisturbed straight trajectories of the accelerating beams and on-axis beam propagating both accelerated and decelerated beams in the linac's SRF structures.

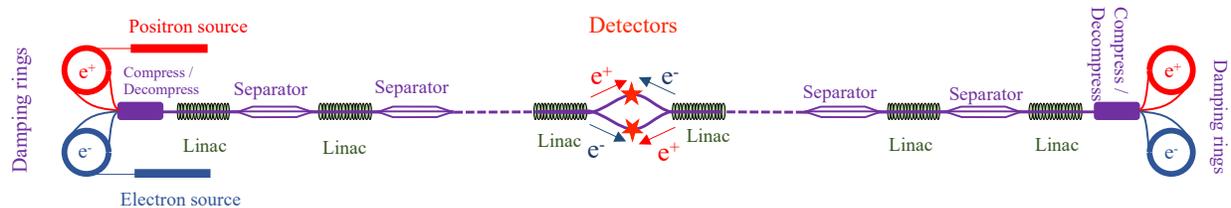

Fig. 1. Schematic of the linear energy recovery e+e- collider with center of mass energy from 90 GeV to 3 TeV. We assume that bunches are compressed 10-fold longitudinally before the acceleration and de-compressed 10-fold before injection in the damping ring.

In contrast with circular e+e- colliders [1-3], ReLiC would collide beams only once with disruption parameter typical for linear colliders [4] to boost the luminosity. ReLiC design practically evades synchrotron radiation losses, which limit average beam currents in circular e+e- colliders. These novel features would allow to operate e+e- collider at c.m. energy from 100 GeV to 3 TeV range, and at luminosity level from $10^{36}$ cm$^{-2}$ sec$^{-1}$ to $10^{37}$ cm$^{-2}$ sec$^{-1}$.

In contrast with traditional e+e- linear colliders [4], where collided beams are dumped at full energy, the ReLiC would recycle both the particles and their energy. These features allow to increase average beams currents and to reach higher luminosity. In addition, in ReLiC we limit the critical energy of beamstrahlung photons to maximum of 250 MeV. While this is needed for re-capturing all collided particles into 2.5 GeV damping rings, it also provides for nearly mono-energetic collisions of highly polarized electron and positron beams.

Fig.2 (a) compares the energy reach and estimated luminosity of ReLiC with other currently proposed e+e- colliders. As can be seen from Fig. 2(b), such collider can cover all process of interest in Higgs sector, including double Higgs and t$\bar{t}$H production and allowing access at tree level to the Higgs self-coupling and top Yukawa coupling. In addition, the 3 TeV c.m. energy reach and the ReLiC's high luminosity opening door for investigating physics beyond standard model.

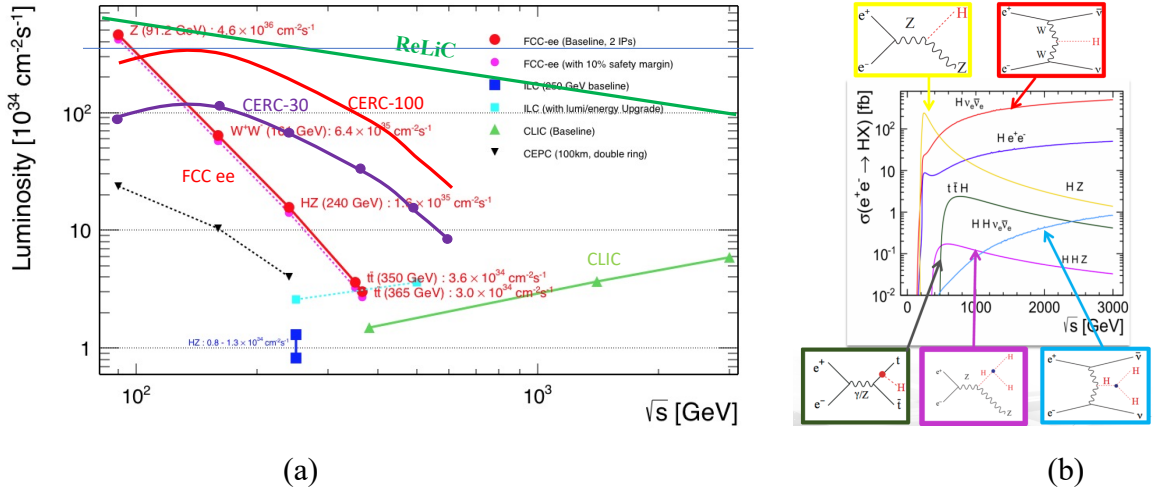

(a)            (b)

Fig. 2. (a) Luminosities for various options for high-energy $e^+e^-$ collider. We added the ReLiC luminosity curve to a plot taken from reference [1]. (b) Cross-sections of various processes in Higgs sector [3].

We used results of our simulations done for the proposed circular ERL-based $e^+e^-$ collider (CERC) [1] to select beam parameters for ReLiC. We evaluated degradation of the beam emittances caused by the beam disruption and transverse offsets in the detectors. We also examined the growth of the energy spread caused by beamstrahlung. In all cases, beam quality remained sufficient for decelerating and injecting into the 2.5 GeV damping rings, located at both sides of the ReLiC. Anticipated particle losses resulting from the burn-off in collisions and scattering on residual gas would be very low and could be easily compensated by a top-off injection from polarized 2.5 GeV injectors.

Used bunches will be kept in damping ring for a time sufficient for restoring their emittance and energy spread to the nominal values before being sent for the next acceleration and collision.

RF power, required to compensate for synchrotron radiation in the damping rings, would be one major component of the overall ReLiC power consumption. In this paper we assumed keeping beams for two e-fold energy damping times in the rings. Logarithmic dependence of the required damping time – and the corresponding power consumption of the damping rings - on the beam's deterioration during collisions provides for robustness of our estimation for ReLiC luminosity and estimated power consumption.

## I. ReLiC concept

As can be seen in Fig.1, electrons and positrons are stored and cooled in damping rings located at both ends of the collider. Bunches of electrons and positrons circulate in damping rings for about two damping times to achieve natural emittances and energy spreads. Short trains of bunches – typically from one to three bunches - are periodically ejected from damping rings and accelerated to the collision energy in SRF linacs. We limit number of bunches per train to reduce energy slew in the bunch train caused by the beam loading to less than $10^{-3}$ of the beam's energy.

We propose to alternate trains of electron and positron bunches in the same linac. Bunches accelerated to the top energy, electron and positron trains are separated to collide in two detectors with bunches emerging from the opposite linac. Using two detectors to collide electron and positron beams propagating in opposite direction, is crucial part of the concept. This allows to use magnetic fields to flat-beam interaction point optics, which is the only viable option for TeV scale colliders. We assume using final focus optics like that proposed for FCCee [2].

After collisions in one of two detectors, beams are decelerated in the opposite SRF linac, and injected into the damping rings located the opposite end of the collider. After cooling in the ring, the same particles would travel in the opposite direction, collide in the other detector and finish where they originated from. Particles lost from the burn-off in collisions or from scattering on residual gas are replaced – topped off – from the injectors.

The other advantage of linear energy recovery system is that accelerated and decelerated beams have the same energy at each point of the collider. This allows to optimize beam focusing in each section of linac and dramatically increase threshold for transverse beam-breakup instability (TBBU). In ReLiC the beams would propagate on axis of SRF linac, i.e. this concept does not require development of a new SRF technology. To avoid parasitic beam collisions outside the detectors, trains of decelerating bunches are separated horizontally from accelerating bunches by periodically placed separators (see Fig. 3).

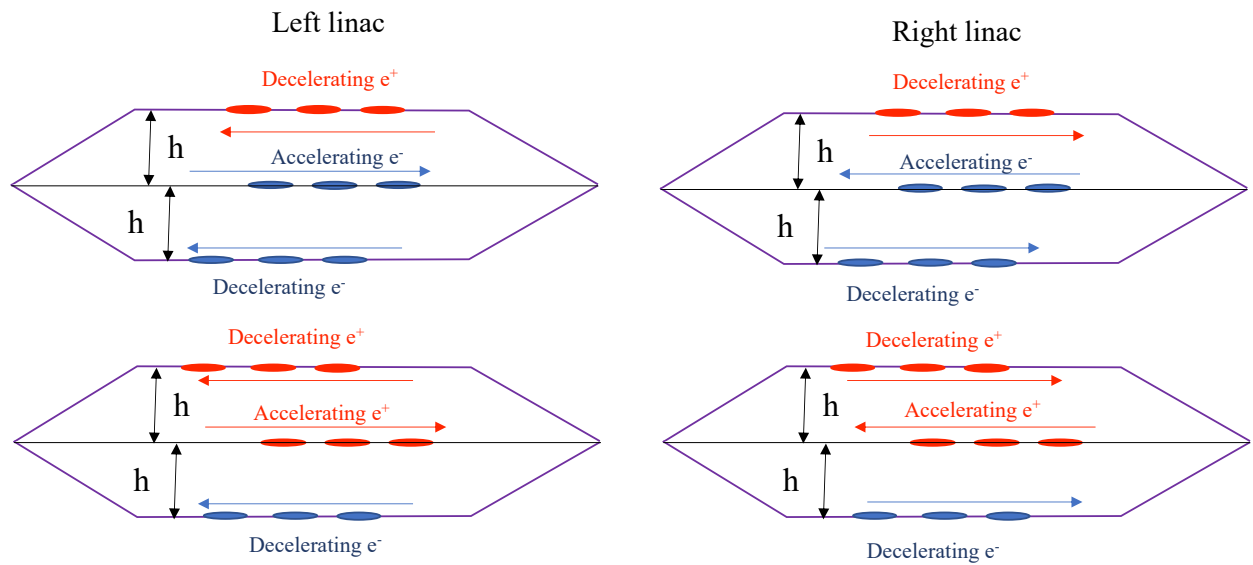

Fig. 3. Separation for trains of electron and positron bunches. Accelerating electrons and positrons beams go straight, while decelerated electrons and positrons are deflected.

Timing of the bunch trains is selected in such way they overlap longitudinally only on separators.

Separators would use combination of DC electric and magnetic fields[1] providing for an undisturbed (straight) trajectory of the accelerated beams. At the same time, decelerating electron and positron bunches will be deflected (naturally, in opposite directions) from the axis of the accelerating beam. After the separator, all particles will return on the linac's axis.

In these limited studies, we assumed that each SRF linac is split in 250-meter sections divided by separators. Length of separators is proportional to the beam energy at their location and was chosen to be 100 m at 1.5 TeV. Separation of the beams is horizontal and is inversely proportional to the square root of the beam energy: beams are separated by at least 10 RMS beam sizes to avoid effects of parasitic beam collisions. In this scenario, the effects from the separators are negligible in terms of synchrotron radiation power, induced energy spread and emittance growth.

In addition, we selected 500 MHz as a convenient and well-established frequency for SRF linac with promises to achieve quality factors exceeding $Q=10^{10}$. We also assumed the use of 5-cell linacs sections as a proven technology with strong higher order modes (HOMs) damping. With a single pass through the linacs, damping of HOMs will ensure that collider can operate at these proposed currents without transverse beam instability (TTBU). We selected the real estate gradient in the linac sections to be at modest level of 12.5 MV/m.

In linear colliders, beam disruption and beamstrahlung effects play important roles in beam dynamics and collision's quality. Destructive power of these effects can be mitigated by using flat colliding beams. In ReLiC we are using ratio of beam horizontal and vertical beam sizes in the IP ranging from 1,000 to 1,500.

Disruption parameters in ReLiC range between 20 and 50 in vertical plane and are negligibly small (~0.01) in horizontal plane. According to simulations [1,3], such collisions will result in 2- to 4-fold increase of the vertical beam emittance. Vertical beam sizes continue to be so minuscule– µm scale to be exact - and do not create any problem with decelerating and transporting used beams into the damping rings.

The most important affect in ReLiC operating very high energies is the beamstrahlung, i.e. emission of high energy photons during beam's collisions. While this is not a challenge at beam energies of 250 GeV or below, at TeV c.m. energies beamstrahlung radiation of high energy photons can create a problem with particle's recovery.

We selected beam parameters and the IP optics for ReLiC in such way that at 3 TeV c.m. energy critical energy of beamstrahlung photons does not exceed 750 MeV with probability of radiating photon with critical energy of 0.27%. In this case, our proposed 10-fold longitudinal decompression of the decelerated beam in combination with 10% acceptance of 2.5 GeV damping ring will be sufficient for practically lossless capture of collided particles. Estimated losses of particles will not exceed one particle per billion.

---

[1] Total Lorentz force $F_x = \pm e\left(E_x + \dfrac{v_z}{c}B_y\right)$ is zero for accelerating particles when $E_x = -\dfrac{v_z}{c}B_y$. Sign of $E_x/B_y$ will depend on direction of accelerating particles, e.g. sign of $v_z$. Decelerated beams have opposite sign of velocity and will be deflected by force of $F_{xdef} = \pm 2eE_x$.

It is worth mentioning that 10% energy acceptance was demonstrated at Duke storage ring. Furthermore, recent results from joined R&D by Cornel University and BNL illustrated that energy acceptance of advanced ring lattice can be significantly larger than 10% [5]. Using such lattice may allow to either reduce required power of synchrotron radiation or, otherwise, to increase ReLiC luminosity.

There is an additional advantage of limiting energy of beamstrahlung photons because it provides for nearly monoenergetic e+e- collisions. This feature is significant advantage of ReLiC operating at TeV c.m. energies, when compared with other designs of linear colliders.

Main ReLiC parameters for two c.m. energies are summarized in Table 1.

Table 1. Key ReLiC parameters for two choices of c.m. energy.

| C.M. energy | GeV | 240 | 3,000 |
|---|---|---|---|
| Length of accelerator | km | 20 | 288 |
| Section length | m | 250 | 250 |
| Bunches per train | | 10 | 21 |
| Particles per bunch | $10^{10}$ | 2.0 | 1.0 |
| Collision frequency | MHz | 12.0 | 25.2 |
| Beam currents in linacs | mA | 38 | 40 |
| $\varepsilon_x$, norm | mm mrad | 4.0 | 4.0 |
| $\varepsilon_y$, norm | µm mrad | 1.0 | 1.0 |
| $\beta_x$ | m | 5 | 100 |
| $\beta_y$, matched | mm | 0.34 | 9.7 |
| $\sigma_z$ | mm | 1 | 17 |
| Disruption parameter, $D_x$ | | 0.01 | 0.002 |
| Disruption parameter, $D_y$ | | 43 | 15 |
| Luminosity per detector | $10^{34}$ cm$^{-2}$sec$^{-1}$ | 172 | 47 |
| Total luminosity | $10^{34}$ cm$^{-2}$sec$^{-1}$ | 343 | 94 |

Parameters for intermediate energies fall between numbers falling between numbers indicated in the table. The critical energy of beamstrahlung photons grow as a high power the beam energy. To contain this growth, it we are increasing both the bunch length and β-functions in the IP to reduce transverse fields generated by colliding beams. This increases result is reduction of the attainable luminosity at higher energies,

Because of the limited efforts devoted to study ReLiC, this set of parameters is not necessarily optimized and further increases of ReLiC luminosity are likely possible.

## II. Performance matrix for ReLiC

| | |
|---|---|
| Attainable c.m. energy : | from 90 GeV to 3 TeV |
| Acceleration rate: | 12.5 MV/m |
| RF power: SRF linear accelerators | 30 MW per each TeV in c.m. |
| | Damping rings, all: ~5 MW per 1 mA |
| Magnet technology: conventional | |
| AC power consumption: | 300 MW for 240 GeV c.m (estimation) |
| | 800 MW for 3 TeV c.m. (estimation) |
| Number of detectors | 2 |
| Luminosity per detector | $1.7 \cdot 10^{36}$ cm$^{-2}$sec$^{-1}$ at 240 GeV c.m. |
| | $4.7 \cdot 10^{35}$ cm$^{-2}$sec$^{-1}$ at 3 TeV c.m. |
| Integrated luminosity per year, fb$^{-1}$ | 6,900 at 240 GeV c.m. |
| | 1,900 at 3 TeV c.m. |
| Final focus | similar to the FCC ee design |
| Normalized beam emittances | horizontal - 4 μm rad, vertical – 1 nm rad |
| Beam generation (preparation) | 2.5 GeV low emittance damping rings |
| Vacuum systems: | conventional for linac, special for damping ring to absorb SR |

*Additional information*:

System includes two 2.5 GeV pulsed electron and positron linacs with very low current to top-off damping rings to compensate for lost particles.

*Staging options*:

As in any linear accelerators, beam energy can be increased by adding accelerating cavities or increasing accelerating gradients. Staging linac tunnel would require relocating the damping rings and injection linacs.

Luminosity upgrades are also possible by increasing the collision rep-rate and RF power in damping rings

Experimental systems upgrades do not depend on ReLiC concept

*Synergies with other concepts and existing facilities:*

Common R&D on high-Q SRF cavities, microphonics suppression systems, HOM dampers.

Common technologies with low emittance light sources, high current e+e- colliders and damping ring lattices with large energy acceptance.

## III. Challenges and discussions

In addition to traditional challenges of linear colliders, ReLiC has one major technical challenge - development of the injection and ejection kickers with MHz rep-rates. Second challenge would be in the damping rings – or probably a series of damping rings - to accommodate large circulating beam currents and absorbing multi-MW SR power. The luminosity is proportional to this power and, if necessary, can be scaled up to $10^{37}$ cm$^{-2}$ sec$^{-1}$ level with natural cost of increased power consumption.

Nevertheless, potential of CW recycling linear collider is its energy and luminosity reach. One can consider reducing collision rate and beam current 10-fold, or relax collision conditions, and still operate with luminosity and energy reach exceeding that of competing concepts. Set of parameters presented in this white paper was not optimized and further improvement are possible.

Needless to say, a detailed technical validation of this concept is needed.